\documentclass[fleqn, 10pt, a4paper]{article}
\usepackage{float, graphics, lineno, color, authblk, multirow, makecell, fancyhdr, lastpage}
\usepackage{newtxtext,newtxmath,amsmath}
\usepackage{orcidlink}
\usepackage[biblabel, nomove]{cite}
\usepackage[left=1.5cm, right=1.5cm, top=2.0cm, bottom=2.0cm, headsep=0.5cm]{geometry}
\usepackage[labelfont={bf,sf}, labelsep=period, justification=raggedright]{caption}

\hypersetup{colorlinks=true, allcolors=blue}
\makeatletter
\renewcommand\normalsize{%
\@setfontsize\normalsize\@xpt\@xiipt
\abovedisplayskip 2.5\p@ \@plus2\p@ \@minus5\p@
\abovedisplayshortskip \z@ \@plus3\p@
\belowdisplayshortskip 6\p@ \@plus3\p@ \@minus3\p@
\belowdisplayskip \abovedisplayskip
\let\@listi\@listI}
\makeatother

\title{\textbf{Heavy-tailed distributions of confirmed COVID-19 cases and deaths in spatiotemporal space}}

\author[1,*]{Peng Liu\orcidlink{0000-0002-9115-2055}}
\author[2,*]{Yanyan Zheng\orcidlink{0000-0002-1092-072X}}
\affil[1]{School of Information, Xi'an University of Finance and Economics, Xi'an 710100, Shaanxi, P.R. China}
\affil[2]{School of Management, Xi'an Polytechnic University, Xi'an 710048, Shaanxi, P.R. China}
\affil[*]{corresponding authors: \href{mailto:pengliu@xaufe.edu.cn}{pengliu@xaufe.edu.cn} (P. Liu), \href{mailto:yanzheng@whu.edu.cn}{yanzheng@whu.edu.cn} (Y. Zheng)}
\date{\today}

\begin{document}

\maketitle
\thispagestyle{empty}

\begin{abstract}
This paper conducts a systematic statistical analysis of the characteristics of the geographical empirical distributions
for the numbers of both cumulative and daily confirmed COVID-19 cases and deaths
at county, city, and state levels over a time span from January 2020 to June 2022.
The mathematical heavy-tailed distributions can be used for fitting the empirical distributions observed in different temporal stages and geographical scales.
The estimations of the shape parameter of the tail distributions using the Generalized Pareto Distribution
also support the observations of the heavy-tailed distributions.
According to the characteristics of the heavy-tailed distributions, the evolution course of the geographical empirical distributions
can be divided into three distinct phases, namely the power-law phase, the lognormal phase I, and the lognormal phase II.
These three phases
could serve as an indicator of the severity degree of the COVID-19 pandemic within an area.
The empirical results suggest important intrinsic dynamics of a human infectious virus spread
in the human interconnected physical complex network.
The findings extend previous empirical studies
and could provide more strict constraints for current mathematical and physical modeling studies,
such as the SIR model and its variants based on the theory of complex networks.
\end{abstract}

{\noindent}\rule[0pt]{18cm}{0.1em}  

\section*{Introduction}
The COVID-19 global pandemic broke out in the human interconnected physical complex network in late 2019.
It has caused unprecedented damage to global public health, society safety, and economy{\cite{WHO-COVID-19, outbreak-1, outbreak-2, name-it, Lancet-death}}.
To control such a pandemic well, 
many countries have taken a variety of containment measures since the beginning phase of this epidemic{\cite{containment_JAMA}}.

Over the past few years,
it has been observed that some countries, states, provinces, cities, and counties
have a huge number of confirmed COVID-19 cases and deaths,
while others have a few confirmed cases and deaths{\cite{WHO-COVID-19, Our-World-in-Data, CSSE-github, CSSE-2}}.
One may intuitively attribute this geographical heterogeneity to the different containment measures adopted in different regions.
It is true that some countries have adopted relatively strict containment measures compared with others.
However, within a country, such as China where unified COVID-19 containment measures are adopted,
similar geographical heterogeneity is still observed.

The huge geographical heterogeneity implies spatial heavy-tailed distribution{\cite{COVID-19-Power-Law, TB-2020, Ahundjanov-2021, Peng-Yanyan}}.
Blasius{\cite{COVID-19-Power-Law}} analyzed data of COVID-19 confirmed cases and deaths at the end of March 2020 for countries worldwide and for counties in the US,
and revealed that the geographical distributions of confirmed cases and deaths follow the truncated power-law.
Similarly, Toda and Beare{\cite{TB-2020}} also investigated the US county-level distribution of confirmed cases during the early phase of this epidemic,
and then found that this distribution obeys the power-law with an exponent close to 1.
Subsequently, Ahundjanov et al.{\cite{Ahundjanov-2021}} computed the Chinese city-level distribution of confirmed cases at the end of May 2020,
and obtained a similar result as that of the US county-level distribution{\cite{Ahundjanov-2021}}.
In 2022, Liu and Zheng{\cite{Peng-Yanyan}} conducted a more comprehensive and systematic analysis
of COVID-19 cumulative and daily confirmed cases and deaths for countries worldwide by considering its temporal evolution characteristics.
This study shows that the power-law and the stretched exponential function can well describe the geographical country-level distributions,
in a time period from the early phase of this epidemic to June 2022{\cite{Peng-Yanyan}}.
In addition to the geographical power-law distribution,
a power-law growth pattern during the early phase of this epidemic was also observed{\cite{Singer-2020, PRE-2020, SR-2021, SK-2022}}.
The power-law pattern reveals important intrinsic dynamics of the COVID-19 spread process in the complex network of human society.

The previous empirical studies focused only on data from the early phase of this epidemic
and did not investigate the differences among different geographical scales.
To obtain a comprehensive understanding of the dynamics of the COVID-19 spread process,
it is valuable to perform a more in-depth and systematic analysis of COVID-19 data across different geographical scales in a time span as long as possible.
Compared with the previous studies{\cite{COVID-19-Power-Law, TB-2020, Ahundjanov-2021, Peng-Yanyan}},
at the county, city, and state/province levels,
this paper extendedly analyzes the COVID-19 time series data from Australia, Canada, China, Denmark, France, Netherlands, New Zealand, the UK, and the US,
and further investigates the spatial and temporal impact on the distributions of both cumulative and daily confirmed cases and deaths.

The heavy-tailed distribution features the concurrence of small and extremely large values.
It has been observed widely in the complex systems of nature and human society{\cite{Power-Law-review1, Power-Law-review2, Power-Law-review3, Power-Law-review4, Power-Law-review5}}.
Typical examples include the distributions of
income and wealth{\cite{weath-dis1, weath-dis2}},
return of financial assets{\cite{return-1, return-2, return-3}},
consumption{\cite{consumption-2017}},
firm size{\cite{firms}},
rainfall depth{\cite{rainfall}},
agricultural land size{\cite{farmlandsize}},
city size{\cite{citysize}},
rank-size of human settlements{\cite{settlement}},
human mobility{\cite{human-mobility-1, human-mobility-2, human-mobility-3, human-mobility-4}},
oil and natural gas production{\cite{oil}},
frequency of words{\cite{words}},
university research activities{\cite{researchactivity}},
degree of complex networks{\cite{BB-1, BB-2, BB-3}}, 
tropical cyclone damages{\cite{Conte-2018}},
COVID-19 coronavirus superspreading{\cite{Wong-2020}},
etc.
Such distribution is an indication of the essential dynamics underlying a complex system,
and often related to the phenomena of self-organized criticality, turbulence, fractals, and so on{\cite{BB-3, criticality, turbulence-1, turbulence-2, fractals}}.
This paper contributes to the literature by presenting the observation of
heavy-tailed distributions in empirical data of confirmed COVID-19 cases and deaths.
This paper also provides another example of the damages from natural disasters characterized by the heavy-tailed distribution{\cite{Conte-2018}}.

To understand the spread dynamics of the COVID-19 pandemic, 
scholars have conducted a large number of theoretical studies using mathematical and physical modeling approaches.
Pullano et al. {\cite{Pullano-2020}} and Gilbert et al.{\cite{Gilbert-Lancet}} estimated the early importation risk by considering air travel flows originating from the infected cities in China.
Jia et al.{\cite{Jia-Nature}} introduced a spatiotemporal risk source model to study the geographical spread and the growth pattern during the early phase based on the population flow data. 
Maier and Brockmann {\cite{Maier-Science}} and Manchein et al.{\cite{Manchein-Chaos}} investigated the impact of containment measures on
the early subexponential and power-law growth patterns.
Arenas et al.{\cite{Arenas-PRX}} introduced an age-stratified mobility-based metapopulation model to study
the spatiotemporal spread by considering mobility and social distancing interventions.
Sardar et al.{\cite{Sardar-CSF}} and Sardar and Rana{\cite{Sardar-2022}} developed mathematical models with lockdown effect
to assess lockdown policy and potential outbreak risks from hospitals and quarantined centers.
Jentsch et al.{\cite{Jentsch-2021}} introduced a coupled social-epidemiological model by considering the interaction between social and epidemiological dynamics.
Musa et al.{\cite{Musa-2021}} introduced a COVID-19 mathematical model which considered the effect of public awareness.
Moore et al.{\cite{Moore-Lancet}} developed a model structured by age and UK region to study the effects of vaccination and non-pharmaceutical interventions.
Özköse et al.{\cite{Ozkose-CSF}} proposed a fractional order model to study the dynamics of the Omicron variant.
Ikram et al.{\cite{Ikram-2022}} developed a stochastic SIVR model for COVID-19.
Sk et al.{\cite{SK-2022}} introduced a completely new model which incorporated the power-law effect of the COVID-19 transmission process.
Chang et al.{\cite{Chang-Nature-2021}} introduced a metapopulation SEIR model using mobile phone data to explain inequities and inform reopening.
Although there are many mathematical and physical modeling studies as stated above,
these studies did not capture the feature of the geographical heavy-tailed distributions presented in some empirical studies{\cite{COVID-19-Power-Law, TB-2020, Ahundjanov-2021, Peng-Yanyan}}.
Blasius{\cite{COVID-19-Power-Law}} proposed a model with two spatial scales to explain the emergence of power-law during the initial phase,
and pointed out that the power-law pattern might be changed in the subsequent phase of this pandemic.
Beare and Toda{\cite{TB-2020}} used a simple mathematical model involving Gibrat's law to explain the emergence of the US county-level power-law distribution.
Ahundjanov et al.{\cite{Ahundjanov-2021}} developed a proportionate random growth model based on Gibrat's law to investigate the emergence of the Chinese city-level power-law distribution.

In this paper, we observe not only the geographical power-law distribution of confirmed COVID-19 cases and deaths in different spatial scales,
but also other heavy-tailed distributions across different temporal phases.
The findings presented in this paper extend previous empirical studies,
and could provide more strict constraints for current mathematical and physical modeling studies,
such as the SIR model and its variants based on the theory of complex networks. 

\section*{Data source}
This paper conducts a detailed analysis of the distributions of the COVID-19 numbers for both cumulative and daily confirmed cases and deaths at county, city, and state/province levels.
The COVID-19 datasets analyzed in this paper are sourced from
the Center for Systems Science and Engineering (CSSE) of Johns Hopkins University{\cite{CSSE-github}}
and
the China Data Lab Dataverse operated by Harvard University{\cite{China-Data-Lab1, China-Data-Lab2}}.
The dataset from CSSE was updated until March 2023,
while the dataset from China Data Lab Dataverse was updated until May 2022.
More details of these datasets can be accessed via the sources listed in Table \ref{table1}
and the scientific articles published in journals of the Lancet Infectious Diseases{\cite{CSSE-2}} and the Data and Information Management{\cite{Harvard-2}}.
Table \ref{table1} lists the sources of the datasets for each geographical scale analyzed in this paper.

\begin{table}[H]
	\centering
	\caption{\textbf{The sources of COVID-19 time series data analyzed in this paper.}}
	\label{table1}
	\begin{tabular}{|c|c|c|}
		\hline
		\textbf{Level}                  &\textbf{Country}                                                                                     &\textbf{Source}\\ \hline
		\makecell[c]{State\\(province)} &\makecell[c]{Australia, Canada, China,\\Denmark, France, Netherlands,\\New Zealand, the UK, the US} &\makecell[c]{Reference \cite{China-Data-Lab1} (for the US) \\ Reference \cite{CSSE-github} (for others)}\\ \hline
		City                            &China                                                                                                &Reference \cite{China-Data-Lab2}\\ \hline
		County                          &The US                                                                                              &Reference \cite{CSSE-github}\\ \hline
	\end{tabular}
\end{table}

\section*{Mathematical heavy-tailed distributions and methods}
Many observables in natural and social systems obey the exponentially bounded distribution.
It decays exponentially ($e^{-x}$) or faster than the exponential (e.g. the normal distribution $e^{-x^{2}/\sigma^{2}}$) as the observable $x$ increases.
It is also known as thin-tailed distribution because of the very small probability of large observable $x$.
However, we also observed extreme values of some observables in complex systems.
These extreme events are important and interesting
since they often play a crucial role in understanding the complex system.
The occurrence of extreme events implies a heavy-tailed distribution that decays more slowly than the exponentially bounded distribution{\cite{networkscience-2016}}.
The heavy-tailed distributions that are often used to describe empirical data include
the power-law (also known as the Pareto distribution or Zipf's law), the truncated power-law, the stretched exponential (also known as the Weibull distribution), and the lognormal distributions{\cite{networkscience-2016}}.

The probability density functions of the power-law, the truncated power-law,
the stretched exponential, and the lognormal distributions
are mathematically written as Eqs (\ref{power-law}) - (\ref{lognormal}), respectively{\cite{networkscience-2016}}.
\begin{eqnarray}
\label{power-law}
p(x) = Cx^{-\alpha};\  \    \       \alpha \in \left(1, +\infty\right),\   x \in \left(0, +\infty \right).
\end{eqnarray}
\begin{eqnarray}
\label{truncated-power-law}
p (x) = Cx^{-\alpha}e^{-\lambda x};\  \  \              \alpha \in \left(1, +\infty \right),\   \lambda \in \left(0, +\infty \right),\   x \in \left(0, +\infty \right).
\end{eqnarray}
\begin{eqnarray}
\label{stretched-exponential}
p (x) = C \left(\lambda x\right)^{\beta - 1} e^{-\left(\lambda x\right)^{\beta}};\  \  \      \lambda \in \left(0, +\infty \right),\   \beta \in \left(0, +\infty \right),\   x \in \left(0, +\infty \right).
\end{eqnarray}
\begin{eqnarray}
\label{lognormal}
p (x) = C \frac{1}{x} e^{- \frac{\left(\ln x - \mu\right)^{2}}{2\sigma^2}};\  \  \    \mu \in \left(-\infty, +\infty \right),\   \sigma \in \left(0, +\infty \right),\   x \in \left(0, +\infty \right).
\end{eqnarray}
For some special cases, we can let support of $x$ is $\left[x_{min},\  +\infty \right)$ or $\left[x_{min},\  x_{max} \right]$ with condition of $x_{min} > 0$.
The $C$ is a normalizer which is different for continuous and discrete random variables.
In our study, we use the discrete form since the numbers of confirmed COVID-19 cases and deaths are integers.

The power-law distribution has been generalized as the Generalized Pareto Distribution (GPD).
The GPD covers both cases of the thin and heavy-tailed distributions.
Therefore, although the GPD might not be suitable for describing an empirical distribution over the entire range of a variable, 
it is often used to estimate the shape of a distribution's tail{\cite{Conte-2018}}.

In this paper,
the power-law, the truncated power-law, the stretched exponential, and the lognormal distributions
are employed to fit the empirical distribution over the observable's entire range
with the popular fitting methods developed by Clauset et al.{\cite{MEJ-2009}} and Klaus et al.{\cite{powerlaw-plosone}}.
The mathematical details of the fitting methods can be found in references{\cite{MEJ-2009, powerlaw-plosone}},
and the algorithm's implementation has been coded in the Python package powerlaw{\cite{powerlaw-package}},
which is widely used in scientific research and thus in this work.
To have a better quantitative understanding of the tail distribution,
this paper also uses the GPD to estimate the tail's shape parameter{\cite{GPD-1, GPD-2}}.

Here we use four statistical hypothesis tests to evaluate the goodness of fit.
The first one is the log-likelihood ratio test.
It can identify which one of the two fits is better.
This method compares one candidate distribution against another,
and calculates two values of $R$ and $p$-value{\cite{powerlaw-package}}.
The positive $R$ indicates that the former is a better fit compared with the latter one, and vice versa {\cite{powerlaw-package}}.
The $p$-value quantifies the significance for that direction,
and that direction is significant when $p$-value is less than 0.05 ( $< 5\%$) {\cite{powerlaw-package}}.
The other tests include
the Kormogorov-Smirnov test (KS test){\cite{Stephens-GOF}},
the Anderson-Darling test (AD test){\cite{Stephens-GOF}},
and the Cramér–von Mises test (CVM test){\cite{Stephens-GOF}}.
The statistics for these three tests are denoted by $D$, $A^{2}$, and $T$ in this paper, respectively.
The $p$-values of these three tests are obtained by performing a Monte Carlo method in our analysis.
It is noteworthy that we should use the KS test with caution, as it is underpowered relative to alternative available test statistics{\cite{powerlaw-package, Stephens-GOF, Lilliefors-KS}}.

\section*{Results and discussion}
\subsection*{The US county-level distributions}
\begin{figure}[H]
	\centering
	\includegraphics[width=1.0\linewidth]{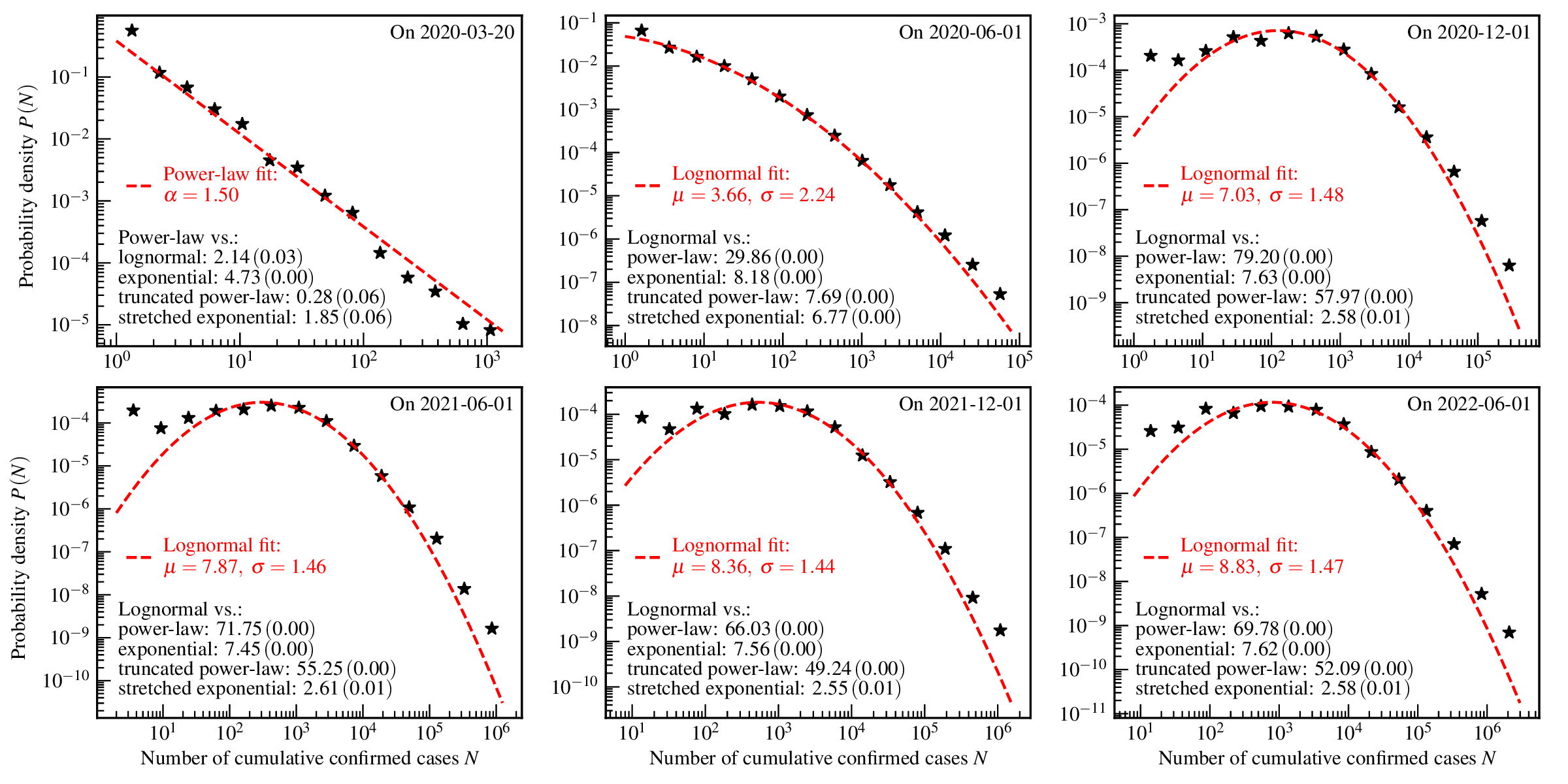}
	\caption{
		\textbf{The US county-level distributions of the numbers of cumulative confirmed COVID-19 cases.}
		The star markers are the empirically estimated probability density $P\left(N\right)$ of a county to have $N$ cumulative confirmed cases by one day. 
		The red dashed curves and legends are the fit results using theoretical distributions.
		The values of $R$ ($p$-value) of the log-likelihood ratio tests are shown as the black text in the bottom left corners of each panel.
	}
	\label{US_TC}
\end{figure}

\begin{figure}[H]
	\centering
	\includegraphics[width=1.0\linewidth]{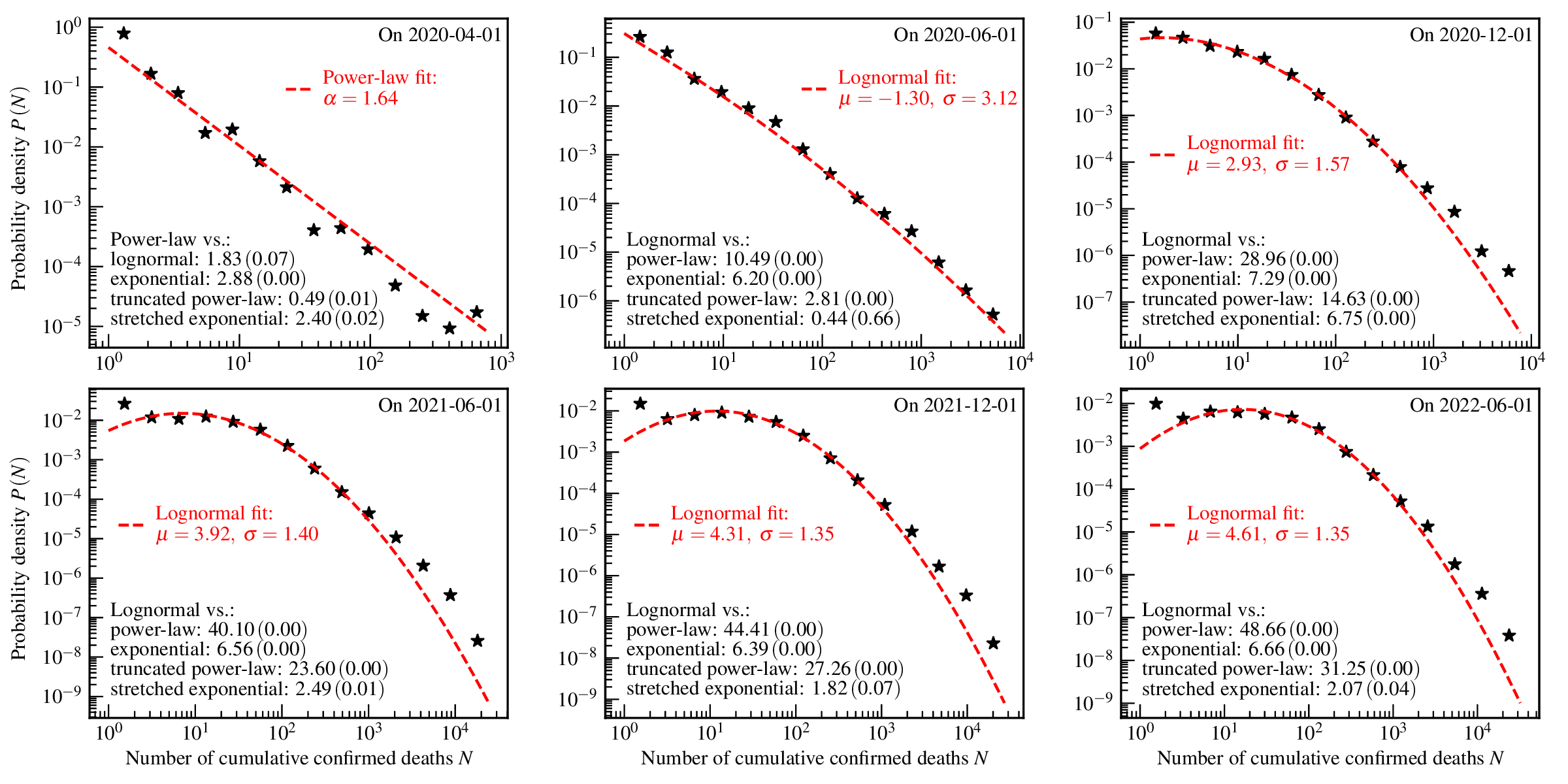}
	\caption{
		\textbf{The US county-level distributions of the numbers of cumulative confirmed COVID-19 deaths.}
		The star markers are the empirically estimated probability $P\left(N\right)$ of a county to have $N$ cumulative confirmed deaths by one day.
		The red dashed curves and legends are the fit results using theoretical distributions.
		The values of $R$ ($p$-value) of the log-likelihood ratio tests are shown as the black text in the bottom left corners of each panel.
	}
	\label{US_TD}
\end{figure}

\begin{figure}[H]
	\centering
	\includegraphics[width=1.0\linewidth]{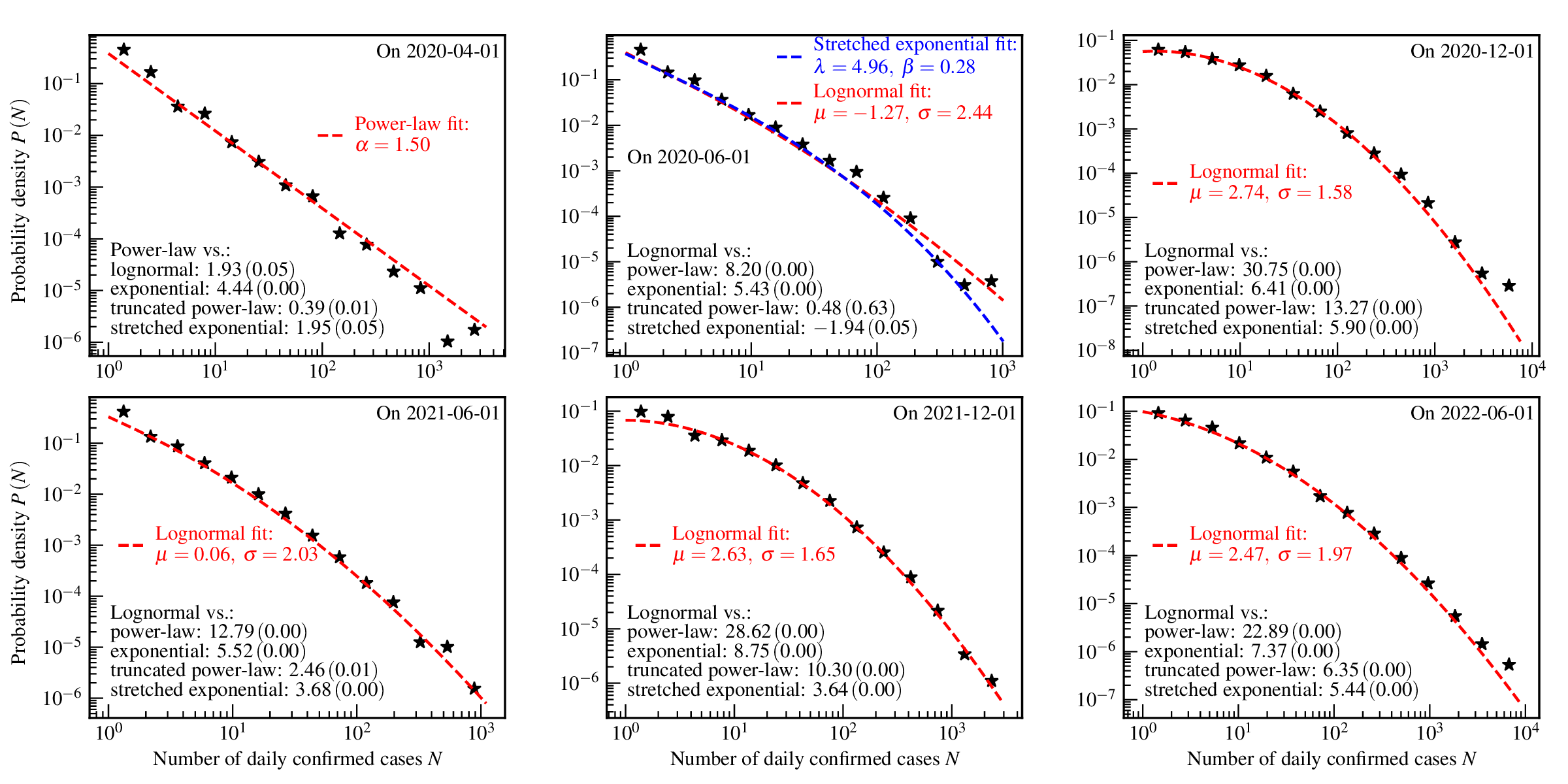}
	\caption{
		\textbf{The US county-level distributions of the numbers of daily confirmed COVID-19 cases.}
		The star markers are the empirically estimated probability density $P\left(N\right)$ of a county to have $N$ daily confirmed cases on one day. 
		The red dashed curves and legends are the results of the lognormal fits.
		For ease of comparison, the results of the stretched exponential fit are presented as the blue dashed curve and legend in the top middle panel.
		The values of $R$ ($p$-value) of the log-likelihood ratio tests are shown as black text in the bottom left corners of each panel.
	}
	\label{US_NC}
\end{figure}

\begin{figure}[H]
	\centering
	\includegraphics[width=1.0\linewidth]{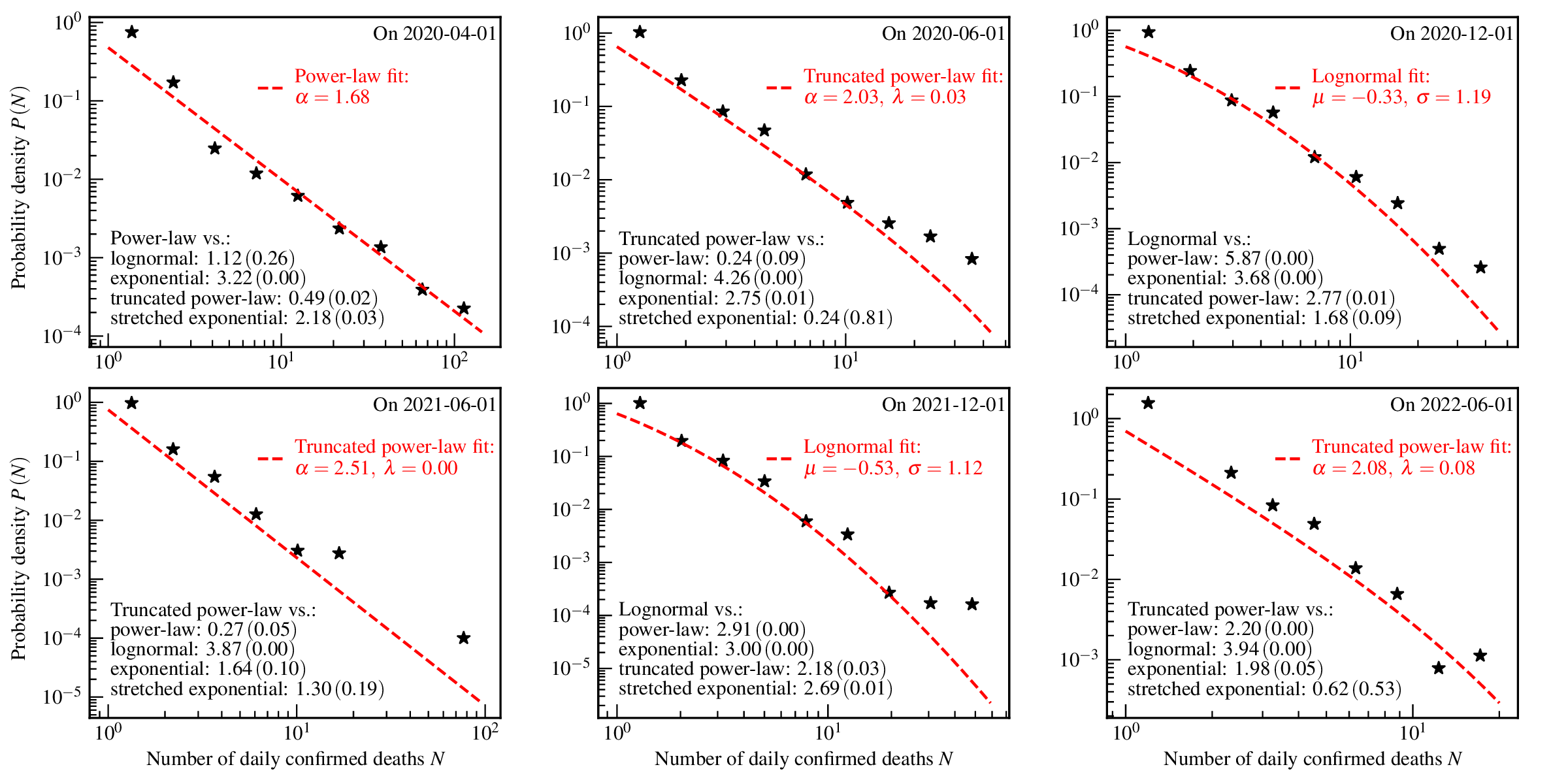}
	\caption{
		\textbf{The US county-level distributions of the numbers of daily confirmed COVID-19 deaths.}
		The star markers are the empirically estimated probability density $P\left(N\right)$ of a county to have $N$ daily confirmed deaths on one day.
		The red dashed curves and legends are the fit results using theoretical distributions.
		The values of $R$ ($p$-value) of the log-likelihood ratio tests are shown as the black text in the bottom left corners of each panel.
	}
	\label{US_ND}
\end{figure}

Here we first analyze the cumulative and daily numbers of confirmed COVID-19 cases and deaths at the US county-level.
Since the COVID-19 pandemic has invaded more than 3,000 counties in the US,
there are enough statistics to investigate the characteristics of the distributions for these four metrics.
Figs \ref{US_TC} to \ref{US_ND} show the evolution behavior of these distributions
over more than two years from the early stages of this pandemic to 1 June 2022.
These four figures use the same techniques of statistical analysis.

Fig \ref{US_TC} shows the US county-level distributions of the numbers of cumulative confirmed cases.
From this figure, it is seen that
the power-law and lognormal distributions describe these empirical data well.
The key parameters of theoretical distributions are obtained by the fitting algorithm{\cite{MEJ-2009, powerlaw-plosone}} implemented in powerlaw package{\cite{powerlaw-package}},
and shown in legends.

The goodness of fits presented in Fig \ref{US_TC} is evaluated by the aforementioned four hypothesis tests.
The black text in the bottom left corners of each panel refers to the log-likelihood ratio test
which compares the exponential,  the power-law, the truncated power-law, the stretched exponential, and the lognormal distributions.
The log-likelihood ratio test indicates that
the lognormal distribution is the best candidate fit for the empirical data presented in the last 5 panels.
For the data presented in the first panel, the power-law might be a better candidate distribution.
However, the log-likelihood ratio test presented in the first panel shows that
the power-law, the truncated power-law, and the stretched exponential distributions
are indistinguishable.
This indistinguishability may be caused by the limitation of statistics in the early stages of the COVID-19 pandemic.
The theoretical distributions,
chosen to fit the empirical data based on the log-likelihood ratio test,
are further tested using the KS test, the AD test, and the CVM test,
whose results are presented in Table \ref{table-USTC}.
From this table, we see that
both the KS test and the CVM test accept these fits at a significance level of 1\%.

\begin{table}[H]
	\centering
	\caption{\textbf{The goodness of fit for the US county-level distributions of the numbers of cumulative confirmed COVID-19 cases.}}
	\label{table-USTC}
	\begin{tabular}{|c|c|c|c|c|}
		\hline
		\textbf{Date}   &\textbf{Distribution}    &\makecell[c]{\textbf{KS test's}\\ \textbf{$D$ ($p$-value)}}  &\makecell[c]{\textbf{AD test's}\\ \textbf{$A^{2}$($p$-value)}}  &\makecell[c]{\textbf{CVM test's} \\ \textbf{$T$ ($p$-value)}}\\ \hline
		2020-03-20 &Power-law     &0.05 (1.00)   &97.30 (1.00)  &18.16 (1.00) \\ \hline
		2020-06-01 &Lognormal     &0.01 (0.88)   &5.27 (0.46)   &0.33 (0.54)  \\ \hline
		2020-12-01 &Lognormal     &0.02 (0.08)   &5.46 (0.01)   &0.60 (0.03)  \\ \hline
		2021-06-01 &Lognormal     &0.02 (0.06)   &7.13 (0.00)   &0.68 (0.02)  \\ \hline
		2021-12-01 &Lognormal     &0.02 (0.10)   &5.07 (0.01)   &0.54 (0.04)  \\ \hline
		2022-06-01 &Lognormal     &0.02 (0.03)   &6.21 (0.00)   &0.69 (0.01)  \\ \hline
	\end{tabular}
\end{table}

Fig \ref{US_TC} exhibits a typical evolution characteristic of the US county-level distribution of cumulative confirmed COVID-19 cases over a time span of more than two years.
In the early stages of this pandemic,  the power-law with an index $\alpha = 1.50$ may be a candidate distribution to describe the empirical data.
A characteristic of such empirical distribution in this phase is that the curve of the probability density function is close to a straight line in a log-log plot.
According to the log-likelihood ratio test,
it is difficult to distinguish the power-law from the truncated power-law and the stretched exponential distributions.
To identify the true distribution pattern underlying empirical data,
the generative mechanism is highly needed to be investigated theoretically by developing mathematical and physical models.

In the subsequent course of this pandemic, the power-law gradually converges to the lognormal distribution,
which is characterized by two key parameters: location $\mu$ and scale $\sigma$.
With the development of this pandemic,
the location $\mu$ gradually increases,
and the scale $\sigma$ decreases except for the data presented in the last panel.
The mode of the lognormal distribution equals $e^{\mu - \sigma^{2}}$.
The fitting result shows that the mode gradually moves to the right along with the evolution course.
Depending on the locational relationship between the mode and the minimum value of the empirical data, 
the subsequent pandemic course has two distinct phases, called lognormal phase I and lognormal phase II.
The mode and the minimum value of the empirical data are close to each other in the lognormal phase I,
and the former is significantly larger than the latter in the lognormal phase II.

Based on the discussion above,
we conclude that the evolution of the empirical distributions over more than two years can be divided into three phases,
namely the power-law phase, the lognormal phase I, and the lognormal phase II.
The distributions of these three phases have the characteristic of the heavy-tailed distribution.
The transformation from the power-law phase to the lognormal phase I is similar to the previous observation on the country-level distributions{\cite{Peng-Yanyan}}.
However, the US county-level distributions especially exhibit the feature of the lognormal phase II, which is not observed in the country-level distributions{\cite{Peng-Yanyan}}.

Fig \ref{US_TD} shows the US county-level distributions of the numbers of cumulative confirmed deaths.
The theoretical distributions,
chosen to fit the empirical data based on the log-likelihood ratio test,
are further tested using the KS test, the AD test, and the CVM test.
The results of these three tests are presented in Table \ref{table-USTD}
and show that these theoretical distributions are acceptable.

A similar evolution process as shown in Fig \ref{US_TC} is observed here.
However, the evolution speed for cumulative confirmed deaths is much slower than that for cumulative confirmed cases.
This phenomenon can be explained by the development of the process of ``infection" events and ``death" events.
The fact that the ``death" event appears later and slower than the ``infection" event causes the slower evolution speed of the distributions of cumulative confirmed deaths.
The effective containment measures contribute partially to this slower evolution.

\begin{table}[H]
	\centering
	\caption{\textbf{The goodness of fit for the US county-level distributions of the numbers of cumulative confirmed COVID-19 deaths.}}
	\label{table-USTD}
	\begin{tabular}{|c|c|c|c|c|}
		\hline
		\textbf{Date}   &\textbf{Distribution}    &\makecell[c]{\textbf{KS test's}\\ \textbf{$D$ ($p$-value)}}  &\makecell[c]{\textbf{AD test's}\\ \textbf{$A^{2}$($p$-value)}}  &\makecell[c]{\textbf{CVM test's} \\ \textbf{$T$ ($p$-value)}}\\ \hline
		2020-04-01 &Power-law     &0.06 (0.99)   &91.67 (1.00)  &18.85 (1.00) \\ \hline
		2020-06-01 &Lognormal     &0.02 (0.59)   &60.05 (0.69)  &8.24 (0.71)  \\ \hline
		2020-12-01 &Lognormal     &0.02 (0.30)   &10.09 (0.09)  &0.94 (0.17)  \\ \hline
		2021-06-01 &Lognormal     &0.02 (0.09)   &6.62 (0.01)   &0.74 (0.03)  \\ \hline
		2021-12-01 &Lognormal     &0.02 (0.16)   &3.99 (0.02)   &0.39 (0.10)  \\ \hline
		2022-06-01 &Lognormal     &0.02 (0.15)   &4.26 (0.01)   &0.43 (0.08)  \\ \hline
	\end{tabular}
\end{table}

For the US county-level distributions of both cumulative confirmed cases and deaths,
the feature of the lognormal phase II is observed.
In accordance with the mathematical model of the lognormal distribution,
one could make a prediction that
if no effective and timely containment measure is taken to control the development of such a pandemic,
all counties will be invaded in the late stages of the COVID-19 course.

Compared with this study,
the previous analysis on the country-level data{\cite{Peng-Yanyan}} does not observe the lognormal phase II at least on 1 June 2022.
This difference between the US county-level and the worldwide country-level may result from the fact that
the severity degree of the COVID-19 pandemic from the perspective of the world is smaller than that from the perspective of the US county.
Actually, these three phases could be an indicator of the severity degree in terms of the geographical scope affected and the number of people infected by the COVID-19 pandemic.

Fig \ref{US_NC} illustrates the US county-level distributions for daily confirmed cases.
The theoretical distributions,
chosen to fit the empirical data based on the log-likelihood ratio test,
are further tested using the KS test, the AD test, and the CVM test.
The results of these three tests are presented in Table \ref{table-USNC}.
The aforementioned four hypothesis tests all indicate that
the power-law and the lognormal distributions are better fits to the empirical data in the early stages and the subsequent course of this pandemic, respectively.
Such observation at the county-level is similar to the former analysis at the country-level \cite{COVID-19-Power-Law, Peng-Yanyan}.
We note that the $R$ for the stretched exponential distribution is negative in the second panel.
Therefore, the stretched exponential distribution is also used to fit that empirical data,
and the other three tests accept this fit.
Only based on the statistical analysis,
we can not distinguish the lognormal distribution from the truncated power-law and the stretched exponential distributions using the empirical data presented in the second panel.
This limitation of the statistical analysis further stimulates the theoretical studies on the generative mechanism using mathematical and physical models.

\begin{table}[H]
	\centering
	\caption{\textbf{The goodness of fit for the US county-level distributions of the numbers of daily confirmed COVID-19 cases.}}
	\label{table-USNC}
	\begin{tabular}{|c|c|c|c|c|}
		\hline
		\textbf{Date}   &\textbf{Distribution}    &\makecell[c]{\textbf{KS test's}\\ \textbf{$D$ ($p$-value)}}  &\makecell[c]{\textbf{AD test's}\\ \textbf{$A^{2}$($p$-value)}}  &\makecell[c]{\textbf{CVM test's} \\ \textbf{$T$ ($p$-value)}}\\ \hline
		2020-04-01                    &Power-law              &0.04 (1.00)   &123.95 (1.00)  &23.25 (1.00)\\ \hline
		{\multirow{2}{*}{2020-06-01}} &Stretched exponential  &0.02 (0.43)   &78.04 (0.69)   &12.52 (0.78)\\ \cline{2-5}
							          &Lognormal              &0.01 (1.00)   &88.06 (0.37)   &14.82 (0.38)\\ \hline
		2020-12-01                    &Lognormal              &0.02 (0.47)   &13.74 (0.08)   &1.34 (0.12) \\ \hline
		2021-06-01                    &Lognormal              &0.01 (0.95)   &70.76 (0.77)   &10.85 (0.77)\\ \hline
		2021-12-01                    &Lognormal              &0.01 (0.99)   &10.19 (0.90)   &0.78 (0.87) \\ \hline
		2022-06-01                    &Lognormal              &0.01 (0.76)   &14.32 (0.27)   &1.29 (0.32) \\ \hline
	\end{tabular}
\end{table}

Fig \ref{US_ND} illustrates the US county-level distributions for daily confirmed deaths.
The KS test, the AD test, and the CVM test presented in Table \ref{table-USND} all show
that the power-law, the truncated power-law, and the lognormal distributions
well describe the empirical data in different stages of the COVID-19 pandemic.
However, we can not make a solid conclusion based on the log-likelihood ratio test due to the limitation of the statistics of the daily data on confirmed deaths.

\newpage

\begin{table}[H]
	\centering
	\caption{\textbf{The goodness of fit for the US county-level distributions of the numbers of daily confirmed COVID-19 deaths.}}
	\label{table-USND}
	\begin{tabular}{|c|c|c|c|c|}
		\hline
		\textbf{Date}   &\textbf{Distribution}    &\makecell[c]{\textbf{KS test's}\\ \textbf{$D$ ($p$-value)}}  &\makecell[c]{\textbf{AD test's}\\ \textbf{$A^{2}$($p$-value)}}  &\makecell[c]{\textbf{CVM test's} \\ \textbf{$T$ ($p$-value)}}\\ \hline
		2020-04-01    &Power-law            &0.08 (0.79)   &38.29 (1.00)    &8.06 (1.00)\\ \hline
		2020-06-01    &Truncated power-law  &0.02 (0.89)   &65.61 (0.31)    &14.58 (0.30)\\ \hline
		2020-12-01    &Lognormal            &0.01 (0.92)   &202.76 (0.41)   &43.61 (0.42)\\ \hline
		2021-06-01    &Truncated power-law  &0.01 (0.97)   &106.16 (0.53)   &24.15 (0.50)\\ \hline
		2021-12-01    &Lognormal            &0.01 (0.72)   &246.66 (0.67)   &54.97 (0.71)\\ \hline
		2022-06-01    &Truncated power-law  &0.01 (0.99)   &106.78 (0.66)   &24.18 (0.63)\\ \hline
	\end{tabular}
\end{table}

\subsection*{The Chinese city-level distributions}
\begin{figure}[H]
	\centering
	\includegraphics[width=1.0\linewidth]{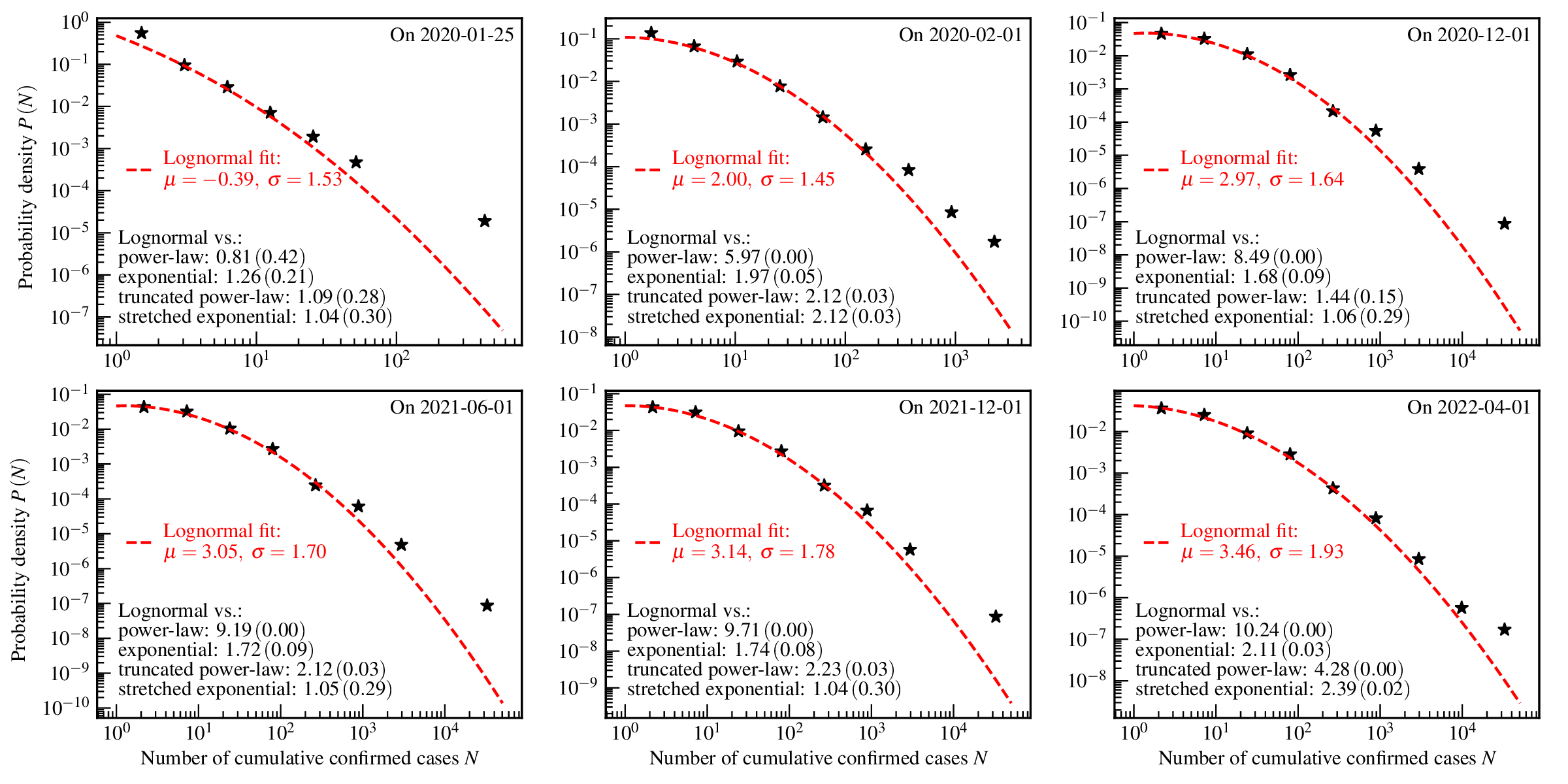}
	\caption{
		\textbf{The Chinese city-level distributions of the numbers of cumulative confirmed COVID-19 cases.}
		The star markers are the empirically estimated probability density $P\left(N\right)$ of a city to have $N$ cumulative confirmed cases by one day.
		The red dashed curves and legends are the results of the lognormal fits.
		The values of $R$ ($p$-value) of the log-likelihood ratio tests are shown as the black text in the bottom left corners of each panel.
	}
	\label{China_TC}
\end{figure}

This paper also analyzes the city-level data of more than 300 Chinese cities.
Fig \ref{China_TC} shows the Chinese city-level distributions of the numbers of cumulative confirmed cases over a time span through the very early stages of this pandemic to 1 April 2022.
There is no illustration for the cumulative confirmed deaths and daily confirmed cases and deaths,
since there are not enough cities with such confirmed cases and deaths.
From the results of the KS test, the AD test, and the CVM test presented in Table \ref{table-ChinaTC}, 
it is evident that the lognormal distribution can well describe the Chinese city-level data in different stages of this pandemic.
However, the log-likelihood ratio test shows
that some theoretical distributions are indistinguishable based on the empirical data on some dates.

Although a solid conclusion is not available based on the log-likelihood ratio test, 
the other hypothesis tests and the shapes of the empirical distributions indicate
that the empirical distributions exhibit the feature of the lognormal phase I in all stages studied here.
In the very early stages of this pandemic, at least by 1 February 2020, the lognormal distribution had been formed.
Compared with the US county-level distributions shown in Fig \ref{US_TC} and the country-level distributions presented in former studies{\cite{COVID-19-Power-Law, Peng-Yanyan}},
the Chinese city-level data formed the lognormal distribution much earlier than the US county-level data and the worldwide country-level data.
The reason for this difference between the Chinese city-level data and
the US county-level data (or the worldwide country-level data) is that the infectious novel coronavirus SARS-CoV-2 first invaded China and then other countries.
Note that no distribution is the form of the lognormal phase II.
This absence of the lognormal phase II reflects
the effect of China's effective pharmaceutical and non-pharmaceutical interventions before mid-2022.

\newpage

\begin{table}[H]
	\centering
	\caption{\textbf{The goodness of fit for the Chinese city-level distributions of the numbers of cumulative confirmed COVID-19 cases.}}
	\label{table-ChinaTC}
	\begin{tabular}{|c|c|c|c|c|}
		\hline
		\textbf{Date}   &\textbf{Distribution}    &\makecell[c]{\textbf{KS test's}\\ \textbf{$D$ ($p$-value)}}  &\makecell[c]{\textbf{AD test's}\\ \textbf{$A^{2}$($p$-value)}}  &\makecell[c]{\textbf{CVM test's} \\ \textbf{$T$ ($p$-value)}}\\ \hline
		2020-01-25  &Lognormal     &0.02 (0.96)     &23.25 (0.33)    &4.59 (0.32)\\ \hline
		2020-02-01  &Lognormal     &0.04 (0.59)     &4.33 (0.25)     &0.49 (0.35)\\ \hline
		2020-12-01  &Lognormal     &0.04 (0.65)     &2.81 (0.09)     &0.27 (0.26)\\ \hline
		2021-06-01  &Lognormal     &0.04 (0.55)     &2.78 (0.09)     &0.27 (0.27)\\ \hline
		2021-12-01  &Lognormal     &0.04 (0.59)     &2.39 (0.13)     &0.24 (0.29)\\ \hline
		2022-04-01  &Lognormal     &0.03 (0.85)     &1.75 (0.25)     &0.17 (0.43)\\ \hline
	\end{tabular}
\end{table}

\subsection*{State/province-level distributions}
\begin{figure}[H]
	\centering
	\includegraphics[width=1.0\linewidth]{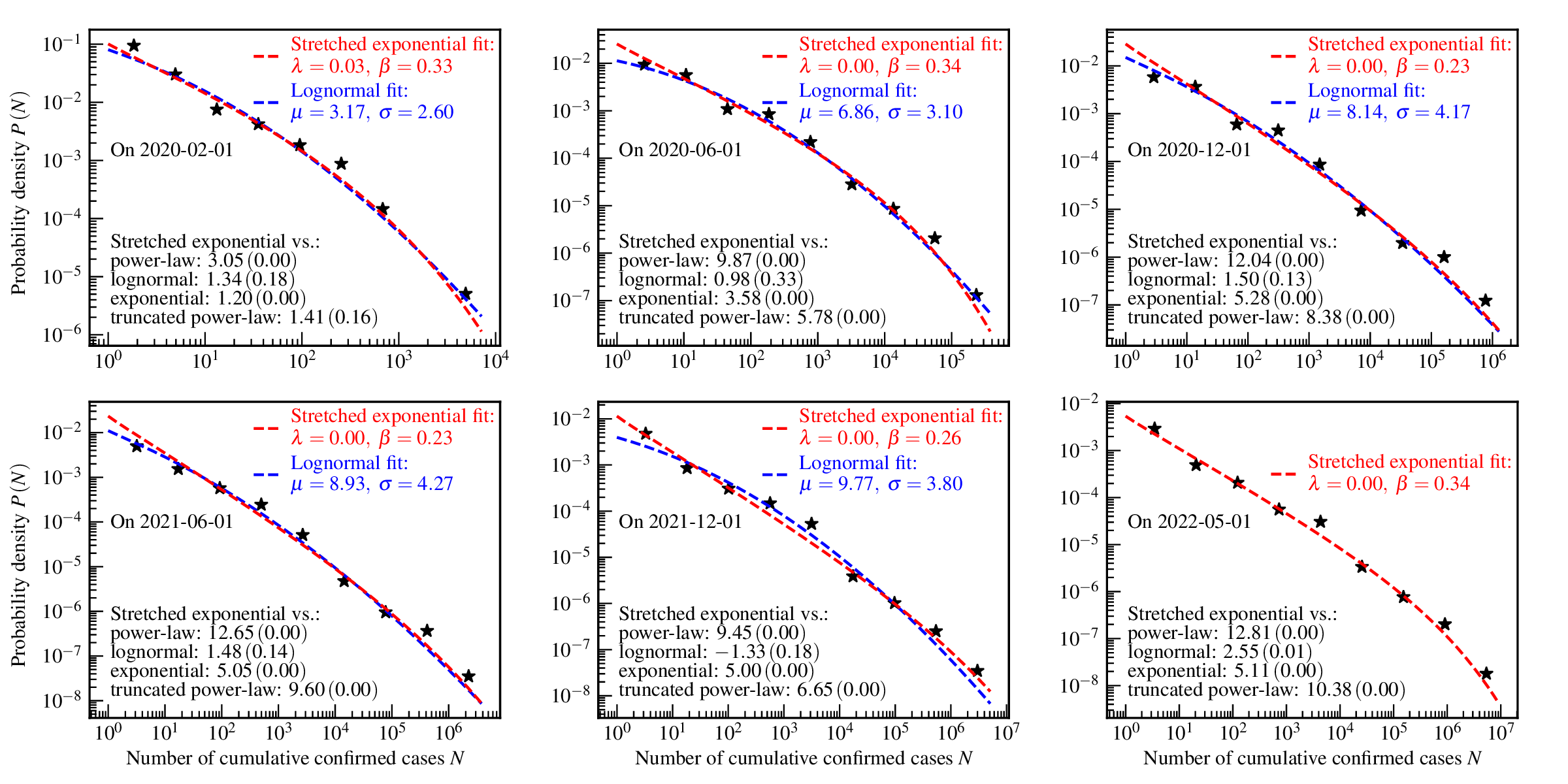}
	\caption{
		\textbf{State/province-level distributions of the numbers of cumulative confirmed COVID-19 cases.}
		The star markers are the empirically estimated probability density $P\left(N\right)$ of a state/province to have $N$ cumulative confirmed cases by one day.
		The red dashed curves and legends are the results of the stretched exponential fits.
		For ease of comparison, the results of the lognormal fits are presented by blue dashed curves and legends.
		The values of $R$ ($p$-value) of the log-likelihood ratio tests are shown as the black text in the bottom left corners of each panel.
	}
	\label{Global_TC}
\end{figure}

\begin{figure}[H]
	\centering
	\includegraphics[width=1.0\linewidth]{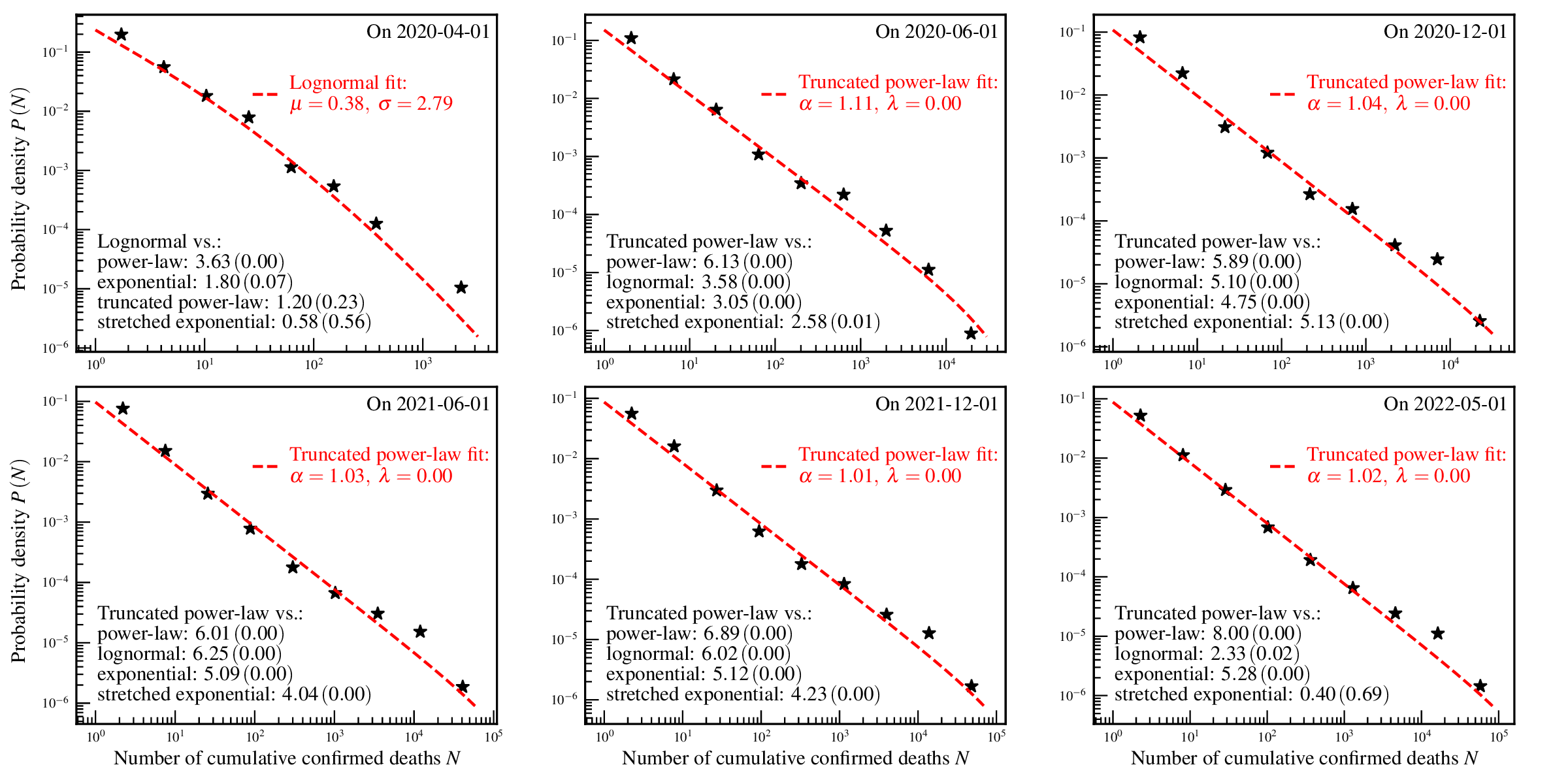}
	\caption{
		\textbf{State/province-level distributions of the numbers of cumulative confirmed COVID-19 deaths.}
		The star markers are the empirically estimated probability density $P\left(N\right)$ of a state/province to have $N$ cumulative confirmed deaths.
		The red dashed curves and legends are the fit results using theoretical distributions.
		The values of $R$ ($p$-value) of the log-likelihood ratio tests are shown as the black text in the bottom left corners of each panel.
		}
	\label{Global_TD}
\end{figure}

In addition to the US county-level and the Chinese city-level data,
this paper also analyzes the state/province-level data
from Australia, Canada, China, Denmark, France, Netherlands, New Zealand, the UK, and the US.
There are more than 100 states and provinces in these 9 countries.
Figs \ref{Global_TC} and \ref{Global_TD} show the state/province-level distributions of
the numbers of cumulative confirmed cases and deaths by considering these more than 100 states and provinces together.
We can not make an illustration of the daily metrics since there are not enough states and provinces with daily confirmed cases and deaths.

For the cumulative confirmed cases,
the state/province-level distribution can be described by the stretched exponential distribution according to the log-likelihood ratio test.
The log-likelihood ratio test also indicates that
the stretched exponential and the lognormal distributions are indistinguishable based on the data presented in the first five panels.
Therefore,  for ease of comparison, we also plot the results of the lognormal fits in the first five panels.
The KS test, AD test, and the CVM test presented in Table \ref{table-GlobalTC} also accept the lognormal fits,
and the stretched exponential fits except for that on 1 December 2021.
It is noteworthy that the empirical distribution shows the feature of lognormal phase I in all stages studied here,
which is similar to the Chinese city-level distribution of cumulative confirmed cases.
As a comparison, the country-level and the US county-level distributions still followed the power-law on March 2020{\cite{COVID-19-Power-Law, Peng-Yanyan}}.

\newpage

\begin{table}[H]
	\centering
	\caption{\textbf{The goodness of fit for the state/province-level distributions of the numbers of cumulative confirmed COVID-19 cases.}}
	\label{table-GlobalTC}
	\begin{tabular}{|c|c|c|c|c|}
		\hline
		\textbf{Date}   &\textbf{Distribution}    &\makecell[c]{\textbf{KS test's}\\ \textbf{$D$ ($p$-value)}}  &\makecell[c]{\textbf{AD test's}\\ \textbf{$A^{2}$($p$-value)}}  &\makecell[c]{\textbf{CVM test's} \\ \textbf{$T$ ($p$-value)}}\\ \hline
		{\multirow{2}{*}{2020-02-01}}   &Stretched exponential     &0.07 (0.93)   &0.61 (0.74)   &0.07 (0.79)\\ \cline{2-5}
						                &Lognormal                 &0.08 (0.85)   &0.73 (0.59)   &0.09 (0.66)\\ \hline
		{\multirow{2}{*}{2020-06-01}}   &Stretched exponential     &0.06 (0.61)   &0.91 (0.41)   &0.14 (0.41)\\ \cline{2-5}
							            &Lognormal                 &0.05 (0.82)   &0.66 (0.59)   &0.09 (0.67)\\ \hline
		{\multirow{2}{*}{2020-12-01}}   &Stretched exponential     &0.08 (0.29)   &2.04 (0.09)   &0.27 (0.17)\\ \cline{2-5}
							            &Lognormal                 &0.08 (0.28)   &1.92 (0.10)   &0.25 (0.20)\\ \hline
		{\multirow{2}{*}{2021-06-01}}   &Stretched exponential     &0.08 (0.31)   &2.15 (0.07)   &0.30 (0.14)\\ \cline{2-5}
							            &Lognormal                 &0.08 (0.32)   &2.03 (0.10)   &0.27 (0.17)\\ \hline
		{\multirow{2}{*}{2021-12-01}}   &Stretched exponential     &0.15 (0.01)   &7.00 (0.00)   &1.20 (0.00)\\ \cline{2-5}
							            &Lognormal                 &0.08 (0.34)   &1.93 (0.10)   &0.28 (0.16)\\ \hline
		{\multirow{2}{*}{2022-05-01}}   &Stretched exponential     &0.06 (0.63)   &1.13 (0.31)   &0.15 (0.40)\\ \cline{2-5}
							            &Lognormal                 &0.08 (0.24)   &2.48 (0.06)   &0.31 (0.13)\\ \hline
	\end{tabular}
\end{table}

\newpage
For the cumulative confirmed deaths, according to the log-likelihood ratio test,
the truncated power-law distribution well describes the data presented in the last five panels,
and the lognormal distribution seems to be a better fit for the data presented in the first panel.
This comparative test also indicates that
the lognormal,  the truncated power-law, and the stretched exponential distributions are indistinguishable based on the data presented in the first panel,
and the truncated power-law and the stretched exponential distributions are indistinguishable based on the data presented in the last panel.
However, all fits are acceptable according to the KS test, AD test, and CVM test which are presented in Table \ref{table-GlobalTD}.
From the shape of these empirical distributions and the fits,
it is evident that these empirical distributions were in the transition stages from the power-law phase to the lognormal phase I. 
At least by 1 May 2022, the data had not yet formed the feature of the lognormal phase II.

\begin{table}[H]
	\centering
	\caption{\textbf{The goodness of fit for the state/province-level distributions of the numbers of cumulative confirmed COVID-19 deaths.}}
	\label{table-GlobalTD}
	\begin{tabular}{|c|c|c|c|c|}
		\hline
		\textbf{Date}   &\textbf{Distribution}    &\makecell[c]{\textbf{KS test's}\\ \textbf{$D$ ($p$-value)}}  &\makecell[c]{\textbf{AD test's}\\ \textbf{$A^{2}$($p$-value)}}  &\makecell[c]{\textbf{CVM test's} \\ \textbf{$T$ ($p$-value)}}\\ \hline
		2020-04-01      &Lognormal                &0.03 (1.00)   &2.28 (0.86)   &0.28 (0.83)\\ \hline
		2020-06-01      &Truncated power-law      &0.05 (0.80)   &1.32 (0.64)   &0.13 (0.72)\\ \hline
		2020-12-01      &Truncated power-law      &0.07 (0.48)   &1.90 (0.17)   &0.22 (0.29)\\ \hline
		2021-06-01      &Truncated power-law      &0.08 (0.35)   &1.99 (0.11)   &0.25 (0.21)\\ \hline
		2021-12-01      &Truncated power-law      &0.07 (0.46)   &1.79 (0.19)   &0.23 (0.26)\\ \hline
		2022-05-01      &Truncated power-law      &0.08 (0.25)   &2.03 (0.10)   &0.26 (0.19)\\ \hline
	\end{tabular}
\end{table}

From the studies discussed above,
it is concluded that
the US county-level, the Chinese city-level, and the state/province-level empirical distributions behave as heavy-tailed distributions.
These heavy-tailed distributions indicate that
a large number of areas are infected slightly
and a few areas are infected tremendously by the COVID-19 pandemic.
The transformation from the power-law phase to the lognormal phase I, and then to the lognormal phase II,
suggests that the geographical heterogeneity is narrowing along with the evolution of this pandemic.

\newpage
\subsection*{Estimations of the shape parameter using the GPD}
To further quantitatively reveal the characteristics of the tails of the US county-level, the Chinese city-level, and the state/province-level empirical distributions,
here we estimate the shape parameters of the tails using the GPD.
Tables \ref{table-GPDUS}, \ref{table-GPDChina}, and \ref{table-GPDGlobal}
report the point and the 95\% confidence interval estimates of the shape parameters over different thresholds, since there is no universal guidance on the selection of threshold.

From Tables \ref{table-GPDUS}, \ref{table-GPDChina}, and \ref{table-GPDGlobal},
we see all point estimates are positive,
which means that these empirical distributions are heavy-tailed.
For a few cases, the 95\% confidence interval estimates show that negative shape parameters are possible,
but there is more possibility that the shape parameters are positive.
Therefore, the GPD estimations of the shape parameters support the conclusion of heavy-tailed distributions,
which is also concluded by fitting the empirical data using the theoretical heavy-tailed distributions.

\begin{table}[H]
	\centering
	\caption{\textbf{Estimation of the shape parameter of the US county-level distribution over thresholds of 50th, 60th, 70th, and 80th percentiles using the GPD.} The numbers in [ ] are the 95\% confidence intervals.}
	\label{table-GPDUS}
	\begin{tabular}{|c|c|c|c|c|}
		\hline
		\textbf{Date}  &\textbf{50th}             &\textbf{60th}             &\textbf{70th}             &\textbf{80th}\\ \hline
		\multicolumn{5}{|c|}{Cumulative confirmed cases}\\ \hline
		2020-03-20     &0.97[0.79, 1.14]          &0.95[0.75, 1.16]          &1.14[0.87, 1.41]          &0.91[0.63, 1.19]\\ \hline
		2020-06-01     &1.23[1.11, 1.34]          &1.18[1.06, 1.31]          &1.13[0.99, 1.26]          &1.09[0.92, 1.26]\\ \hline
		2020-12-01     &0.88[0.79, 0.98]          &0.89[0.78, 1.00]          &0.80[0.67, 0.92]          &0.62[0.49, 0.76]\\ \hline
		2021-06-01     &0.94[0.84, 1.04]          &0.92[0.81, 1.04]          &0.77[0.65, 0.89]          &0.65[0.51, 0.78]\\ \hline
		2021-12-01     &0.85[0.76, 0.95]          &0.84[0.73, 0.94]          &0.70[0.58, 0.81]          &0.56[0.44, 0.69]\\ \hline
		2022-06-01     &0.92[0.82, 1.02]          &0.88[0.77, 0.99]          &0.75[0.63, 0.86]          &0.60[0.47, 0.73]\\ \hline
		\multicolumn{5}{|c|}{Cumulative confirmed deaths}\\ \hline
		2020-04-01     &0.84[0.61, 1.06]          &0.84[0.61, 1.06]          &0.88[0.62, 1.15]          &0.96[0.63, 1.29]\\ \hline
		2020-06-01     &1.10[0.96, 1.24]          &1.13[0.98, 1.29]          &1.14[0.96, 1.32]          &1.21[0.97, 1.45]\\ \hline
		2020-12-01     &0.91[0.81, 1.01]          &0.92[0.81, 1.03]          &0.91[0.79, 1.04]          &0.84[0.69, 0.99]\\ \hline
		2021-06-01     &0.88[0.79, 0.98]          &0.95[0.83, 1.06]          &0.80[0.68, 0.92]          &0.77[0.63, 0.92]\\ \hline
		2021-12-01     &0.81[0.72, 0.91]          &0.81[0.70, 0.92]          &0.71[0.60, 0.82]          &0.65[0.52, 0.78]\\ \hline
		2022-06-01     &0.82[0.72, 0.91]          &0.79[0.69, 0.90]          &0.69[0.58, 0.80]          &0.62[0.49, 0.76]\\ \hline
		\multicolumn{5}{|c|}{Daily confirmed cases}\\ \hline
		2020-04-01     &1.10[0.93, 1.26]          &1.11[0.91, 1.30]          &1.16[0.93, 1.39]          &1.05[0.78, 1.31]\\ \hline
		2020-06-01     &0.75[0.61, 0.89]          &0.69[0.54, 0.83]          &0.64[0.47, 0.81]          &0.51[0.32, 0.69]\\ \hline
		2020-12-01     &0.82[0.72, 0.92]          &0.77[0.66, 0.87]          &0.74[0.62, 0.87]          &0.61[0.47, 0.75]\\ \hline
		2021-06-01     &0.63[0.52, 0.74]          &0.64[0.51, 0.77]          &0.66[0.50, 0.81]          &0.62[0.44, 0.81]\\ \hline
		2021-12-01     &0.64[0.55, 0.73]          &0.63[0.53, 0.72]          &0.59[0.48, 0.70]          &0.57[0.43, 0.71]\\ \hline
		2022-06-01     &0.99[0.87, 1.11]          &0.97[0.83, 1.11]          &0.90[0.74, 1.05]          &0.73[0.56, 0.89]\\ \hline
		\multicolumn{5}{|c|}{Daily confirmed deaths}\\ \hline
		2020-04-01     &0.78[0.45, 1.12]          &0.83[0.38, 1.27]          &0.55[0.09, 1.01]          &0.47[0.00, 0.93]\\ \hline
		2020-06-01     &0.37[0.15, 0.58]          &0.44[0.12, 0.76]          &0.44[0.12, 0.76]          &0.41[-0.05, 0.87]\\\hline
		2020-12-01     &0.20[0.08, 0.31]          &0.20[0.08, 0.31]          &0.22[0.06, 0.38]          &0.22[0.06, 0.38]\\ \hline
		2021-06-01     &0.36[0.14, 0.58]          &0.36[0.14, 0.58]          &0.40[0.07, 0.73]          &0.40[0.07, 0.73]\\ \hline
		2021-12-01     &0.19[0.11, 0.28]          &0.29[0.15, 0.44]          &0.29[0.15, 0.44]          &0.31[0.10, 0.52]\\ \hline
		2022-06-01     &0.03[-0.12, 0.17]         &0.03[-0.12, 0.17]         &0.04[-0.17, 0.24]         &0.02[-0.26, 0.30]\\ \hline
	\end{tabular}
\end{table}

\begin{table}[H]
	\centering
	\caption{\textbf{Estimation of the shape parameter of the Chinese city-level distribution over thresholds of 50th, 60th, 70th, and 80th percentiles using the GPD.} The numbers in [ ] are the 95\% confidence intervals.}
	\label{table-GPDChina}
	\begin{tabular}{|c|c|c|c|c|}
		\hline
		\textbf{Date}  &\textbf{50th}             &\textbf{60th}             &\textbf{70th}             &\textbf{80th}\\ \hline
		\multicolumn{5}{|c|}{Cumulative confirmed cases}\\ \hline
		2020-01-25     &0.64[0.34, 0.95]          &0.73[0.34, 1.13]          &0.81[0.32, 1.30]          &0.84[0.25, 1.43]\\ \hline
		2020-02-01     &0.96[0.66, 1.27]          &1.06[0.70, 1.43]          &1.08[0.64, 1.51]          &0.99[0.44, 1.55]\\ \hline
		2020-12-01     &1.13[0.83, 1.44]          &1.38[0.97, 1.78]          &1.38[0.89, 1.86]          &1.29[0.69, 1.90]\\ \hline
		2021-06-01     &1.19[0.87, 1.52]          &1.46[1.02, 1.90]          &1.31[0.83, 1.78]          &1.20[0.61, 1.80]\\ \hline
		2021-12-01     &1.25[0.90, 1.59]          &1.35[0.92, 1.78]          &1.21[0.73, 1.68]          &0.91[0.42, 1.39]\\ \hline
		2022-04-01     &1.33[0.96, 1.71]          &1.28[0.85, 1.71]          &1.03[0.61, 1.44]          &0.95[0.49, 1.41]\\ \hline
	\end{tabular}
\end{table}

\newpage
\begin{table}[H]
	\centering
	\caption{\textbf{Estimation of the shape parameter of the state/province-level distribution over thresholds of 50th, 60th, 70th, and 80th percentiles using the GPD.} The numbers in [ ] are the 95\% confidence intervals.}
	\label{table-GPDGlobal}
	\begin{tabular}{|c|c|c|c|c|}
		\hline
		\textbf{Date}  &\textbf{50th}             &\textbf{60th}             &\textbf{70th}             &\textbf{80th}\\ \hline
		\multicolumn{5}{|c|}{Cumulative confirmed cases}\\ \hline
		2020-02-01     &0.67[0.07, 1.28]          &0.87[0.02, 1.73]          &0.90[-0.01, 1.82]         &1.09[-0.20, 2.38]\\ \hline
		2020-06-01     &0.99[0.34, 1.64]          &0.50[0.06, 0.95]          &0.39[-0.05, 0.84]         &0.37[-0.15, 0.90]\\ \hline
		2020-12-01     &0.37[-0.04, 0.77]         &0.13[-0.17, 0.44]         &0.10[-0.23, 0.43]         &0.28[-0.24, 0.80]\\ \hline
		2021-06-01     &0.45[0.02, 0.88]          &0.22[-0.11, 0.55]         &0.14[-0.19, 0.48]         &0.41[-0.18, 0.99]\\ \hline
		2021-12-01     &0.37[-0.01, 0.75]         &0.20[-0.12, 0.53]         &0.20[-0.17, 0.56]         &0.25[-0.23, 0.73]\\ \hline
		2022-05-01     &0.25[-0.06, 0.56]         &0.16[-0.13, 0.46]         &0.18[-0.16, 0.51]         &0.47[-0.11, 1.05]\\ \hline
		\multicolumn{5}{|c|}{Cumulative confirmed deaths}\\ \hline
		2020-04-01     &1.27[0.62, 1.93]          &1.03[0.38, 1.68]          &1.25[0.32, 2.19]          &0.72[0.07, 1.37]\\ \hline
		2020-06-01     &0.92[0.34, 1.51]          &0.67[0.12, 1.22]          &0.53[-0.05, 1.11]         &0.33[-0.17, 0.83]\\ \hline
		2020-12-01     &0.37[-0.03, 0.77]         &0.28[-0.13, 0.70]         &0.17[-0.24, 0.57]         &0.46[-0.35, 1.27]\\ \hline
		2021-06-01     &0.34[-0.04, 0.73]         &0.25[-0.14, 0.64]         &0.23[-0.20, 0.66]         &0.24[-0.36, 0.84]\\ \hline
		2021-12-01     &0.38[-0.03, 0.78]         &0.20[-0.15, 0.55]         &0.17[-0.23, 0.56]         &0.28[-0.33, 0.88]\\ \hline
		2022-05-01     &0.35[-0.04, 0.73]         &0.21[-0.16, 0.58]         &0.16[-0.24, 0.56]         &0.22[-0.37, 0.81]\\ \hline
	\end{tabular}
\end{table}

\section*{Conclusions}
The COVID-19 global pandemic has caused unprecedented damage to global public health, society safety, and the economy.
Understanding the spread dynamics of this pandemic is a challenging task.
The geographical distribution characteristics of the numbers of confirmed COVID-19 cases and deaths
suggest important intrinsic dynamics underlying a physical complex network
in which the human infectious virus spreads.
To investigate the distributions in different stages of this pandemic and different geographical scales of the pandemic spread,
at the county, city,  and state/province levels,
this paper systematically analyzes the time series data of the confirmed COVID-19 cases and deaths
from the early stages of this pandemic to June 2022.

The distributions,
for both cumulative and daily confirmed cases and deaths at the US county-level,
for cumulative confirmed cases at the Chinese city-level, 
and for both cumulative confirmed cases and deaths at the state/province-level of Australia, Canada, China, Denmark, France, Netherlands, New Zealand, the UK, and the US,
are statistically analyzed.
The statistical analysis provides
evidence that these geographical distributions can be described by the heavy-tailed distributions in different stages of this pandemic.
These heavy-tailed distributions indicate that
a large number of areas are infected slightly
and a few areas are infected tremendously by the COVID-19 pandemic.
The evolution of this pandemic can be divided into three distinct phases, namely the power-law phase,
the lognormal phase I, and the lognormal phase II.
The distributions in different phases have different shapes.
The shape of the distribution in the power-law phase is close to a straight line in a log-log plot.
For the lognormal phase I and phase II,
the shape of a distribution is a concave curve in a log-log plot.
The difference between these two lognormal phases is the locational relationship between the mode and the observed minimum value.
The location of the mode is close to the observed minimum value for the lognormal phase I,
and the former one is significantly larger than the latter one for the lognormal phase II.

The US county-level cumulative distributions feature these three phases.
In the early stages, the distributions have the feature of the power-law phase.
Then the distributions gradually converge to the lognormal phase I and phase II.
The evolution speed for the cumulative confirmed deaths is slower than that for the cumulative confirmed cases
because of the difference between the dynamics of the ``death" event and the ``infection" event.
The Chinese city-level distributions for the cumulative confirmed cases
only have the feature of the lognormal phase I across the entire stages we studied in this paper.
As a comparison with the US county-level distributions,
the absence of the power-law phase and the lognormal phase II indicates
that the evolution speed for the Chinese city-level distribution is faster in the early stages of this pandemic,
and that speed is slower in the middle and late stages.
The reason for this difference in evolution speed between the Chinese city-level data and
the US county-level data is that the infectious novel coronavirus first invaded China and then other countries,
and the further development of this pandemic
was significantly suppressed by the effective pharmaceutical and non-pharmaceutical interventions taken in China.
The state/province-level data of cumulative confirmed cases have a similar evolution process of the Chinese city-level data,
and that data for cumulative confirmed deaths show a much slower speed of evolution compared with the Chinese city-level data.

These three phases could be an indicator of
the severity degree of the COVID-19 pandemic within an area.
Generally, the power-law phase, 
the lognormal phase I,
and the lognormal phase II 
represent low, middle, and high severity degrees of the COVID-19 pandemic within an area, respectively.
Whether the COVID-19 number in an area forms the distribution following these three phases,
depends on the evolution speed and the severity degree of the COVID-19 pandemic in that area.
The transformation from the power-law phase to the lognormal phase I, and then to the lognormal phase II,
suggests that the geographical heterogeneity is narrowing along with the evolution of this pandemic.

The findings presented in this paper extend previous empirical studies.
And it provides another example of the damages from natural disasters characterized by the heavy-tailed distribution.
This study could provide more strict constraints for current mathematical and physical modeling studies,
such as the SIR model and its variants based on the theory of complex networks.
Such a study is also especially important for predicting the COVID-19 evolution
as many countries are stopping tracking COVID-19\cite{STOPING, STOPING2}.

\section*{Acknowledgements}
This work was supported by the Humanities and Social Sciences Youth Foundation of Ministry of Education of China (Contract No. 22YJCZH107) and
the Shaanxi Science and Technology Department, P.R. China (Contract No. 2023-JC-QN-0093).

\section*{Author contributions}
Peng Liu and Yanyan Zheng made important contributions to this publication in the acquisition of data and data analysis.
Peng Liu wrote the manuscript.
All authors reviewed and approved the submitted manuscript.

\section*{Competing interests}
The authors have declared that no competing interests exist.

\section*{Data Availability Statement}
All data for this study have been deposited in public data repositories GitHub and Harvard Dataverse. The links to datasets are listed here:
\url{https://github.com/CSSEGISandData/COVID-19/tree/master/csse_covid_19_data/csse_covid_19_time_series},
\url{https://doi.org/10.7910/DVN/MR5IJN}, and
\url{https://doi.org/10.7910/DVN/HIDLTK}.
The code used in this study is the Python package powerlaw. This package has been deposited in GitHub: \url{https://github.com/jeffalstott/powerlaw}.

\footnotesize
{
	\bibliography{refs}
}

\end{document}